# Determination of incommensurate modulated structure in Bi$_2$Sr$_{1.6}$La$_{0.4}$CuO$_{6+\delta}$ by aberration-corrected transmission electron microscopy

Binghui Ge[1, a], Yumei Wang[1], Fanghua Li[1], Huiqian Luo[1], Haihu Wen[1, b]
Rong Yu[2], Zhiying Cheng[2], Jing Zhu[2]

[1]*Beijing National Laboratory for Condensed Matter Physics,*
*Institute of Physics, Chinese Academy of Sciences, Beijing 100190, China*
[2]*Beijing National Center for Electron Microscopy, Tsinghua University, Beijing 100084, China*

## Abstract

Incommensurate modulated structure (IMS) in Bi$_2$Sr$_{1.6}$La$_{0.4}$CuO$_{6+\delta}$ (BSLCO) has been studied by aberration corrected transmission electron microscopy in combination with high-dimensional (HD) space description. Two images in the negative $C$s imaging (NCSI) and passive $C$s imaging (PCSI) modes were deconvoluted, respectively. Similar results as to IMS have been obtained from two corresponding projected potential maps (PPMs), but meanwhile the size of dots representing atoms in the NCSI PPM is found to be smaller than that in PCSI one. Considering that size is one of influencing factors of precision, modulation functions for all unoverlapped atoms in BSLCO were determined based on the PPM obtained from the NCSI image in combination with HD space description.

## Introduction

Incommensurate modulated structure (IMS) has been found to be strongly correlated with the superconductivity in bismuth system cuprate superconductors[1]. To understand their relationship and consequently the mechanism of superconductivity, quantitive structural determination of IMS is needed [2]. High-resolution transmission electron microscopy (HRTEM) is a powerful method for structural determination, especially for minute crystals or polycrystals. But due to existence of objective lens aberrations a random high-resolution (HR) image may not reflect the real structure. Even if at the Scherzer focus condition [3], point resolution of conventional medium-voltage electron microscopes is limited at about 0.2 nm such that not all atoms can be resolved. Posterior image processing [4-14] was usually used

---
[a] Corresponding author. Email: bhge@iphy.ac.cn
[b] Present address: Department of Physics, Nanjing University, Nanjing, 210093, China.

to correct the image distortion and enhance the resolution to the information limit.

Great progress has been made since aberration correctors was invented in 1990s [15-18] that point resolution is enhanced even beyond the information limit such that light atoms such as oxygen in oxides can be directly observed [19]. As the resolution of microscope is enhanced to the subangstrom scale, however, image contrast becomes more sensitive to changes of imaging condition [20-22]. Even though at the Scherzer focus condition, artifacts may be introduced into the image when spherical aberration is corrected to close to 0. Therefore, image processing is still necessary for structural determination in aberration-corrected electron microscopy [20-22].

In this article, the method of image deconvolution is applied to study IMS in optimally doped $Bi_2Sr_{1.6}La_{0.4}CuO_{6+\delta}$ (BSLCO). Two high-resolution (HR) images with different imaging mode, positive $C$s imaging (PCSI) and negative $C$s imaging (NCSI) [19, 23-26], are analyzed and compared. From the projected potential map (PPM) of IMS obtained after deconvolution the modulation functions for all unoverlapped atoms including Cu and O in $CuO_2$ planes are determined with the help of the high-dimensional (HD) space description [27].

## 2. HD space description [27]

Due to absence of three-dimensional translation symmetry, IMS cannot be described by three-dimensional crystallography. HD space description [27] should be taken into consideration during structural determination of IMS, and will be introduced briefly in the following.

In IMS, there are two kinds of reflections in the reciprocal space. One is relatively stronger, namely main reflection, which reflects the information of the average structure; the other is weaker, namely satellite reflection, which carries the modulated information. To index all reflections, extra modulation wave vector $q$ should be introduced. For one-dimensional (1D) IMS reciprocal lattice vectors are expressed as following,

$$H = h a^* + k b^* + l c^* + mq, \qquad (1)$$

where $h$, $k$, $l$ and $m$ are integers, $a^*$, $b^*$ and $c^*$ the reciprocal basis vectors, and the modulation wave vector is expressed as $q = \alpha a^* + \beta b^* + \gamma c^*$ with at least one of $\alpha$, $\beta$ and $\gamma$ being irrational.

One vector $e$ perpendicular to the space ($a$, $b$, $c$) is introduced to construct a four-dimensional (4D) periodic structure ($a - \alpha e, b - \beta e, c - \gamma e, e$), in which atoms is one dimensional and periodic along the fourth axis $e$ with different shape for different kinds of modulations. For instance, atom looks like a wavy string with uniform thickness in displacive modulations and like a straight string with a fluctuating thickness in occupational modulations.

## 3. Experimental

Single crystal BSLCO samples with high quality were prepared from mixed powders of $Bi_2O_3$, $SrCO_3$, $La_2O_3$ and CuO with nominal compositions by the traveling-solvent floating-zone technique, and post-annealing experiment was carried out in flowing oxygen in high temperature [28]. The [100] specimen for electron microscope observation was prepared by cutting, mechanical polishing and ion milling, as described in Ref. [1]. The specimen was observed using a Titan 80-300 electron microscope with the information limit about 0.8 Å. The image processing such as superspace symmetry average and image deconvolution was performed by using VEC [29].

## 4. Results and discussion

### 4.1 Determination of PPM of IMS

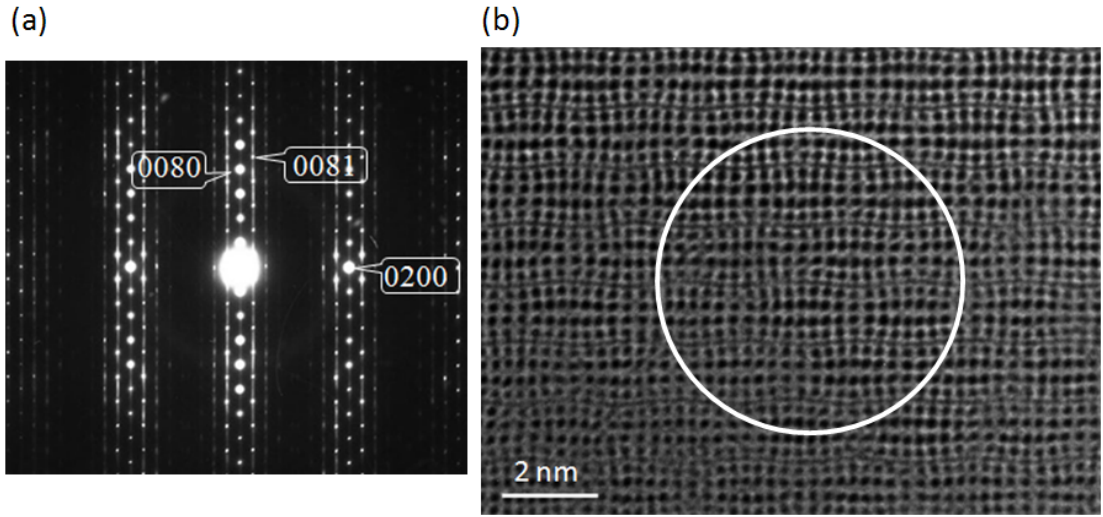

Fig.1 (a) and (b) [100] electron diffraction pattern and HR image with $Cs$ = 50 μm, respectively. The circular area in (b) was selected for further image processing

Same as in its mother system $Bi_2Sr_2CuO_{6+\delta}$ [1], there exists 1D IMS in **bc** plane of BSLCO. The modulated wave vector is determined to be $\boldsymbol{q} = 0.24\boldsymbol{b}^* + \boldsymbol{c}^*$ from its [100] projected electron diffraction pattern as shown in Fig. 1(a) with $\boldsymbol{b}^*$ and $\boldsymbol{c}^*$ the reciprocal basis vectors of BSLCO, and its superspace group $Bbmb(0\beta 1)$ [30]. Fig. 1(b) is a corresponding HR image with residual aberrations $Cs$ = 0.05 mm and $C1$ -11 nm. After Fourier filtering to reduce the random noise on the selected circular area as shown in Fig. 1(b) HD symmetry averaging was carried out with superspace group $Bbmb(0\beta 1)$ to correct the influence of crystal tilt and/or beam tilt, and also correct residual azimuth-dependent aberrations to some extent, say astigmatism and coma. For the image is taken near the Scherzer focus, two adjacent big black dots in the symmetry-averaged image (see Fig. 2(a)) represent Bi(O) atomic columns according the basic structure model of $Bi_2Sr_2CuO_6$ (BSCO) as shown in Fig. 2(b) [31], and small black dots Sr(La, O) columns. Changes of contrast and atomic positions in Bi(O) and Sr(La, O) layers confirm the existence of both compositional and displacive modulations, which seems in the form of harmonic waves. Due to the residual aberrations Cu and O columns in the $CuO_2$ plane are not resolved.

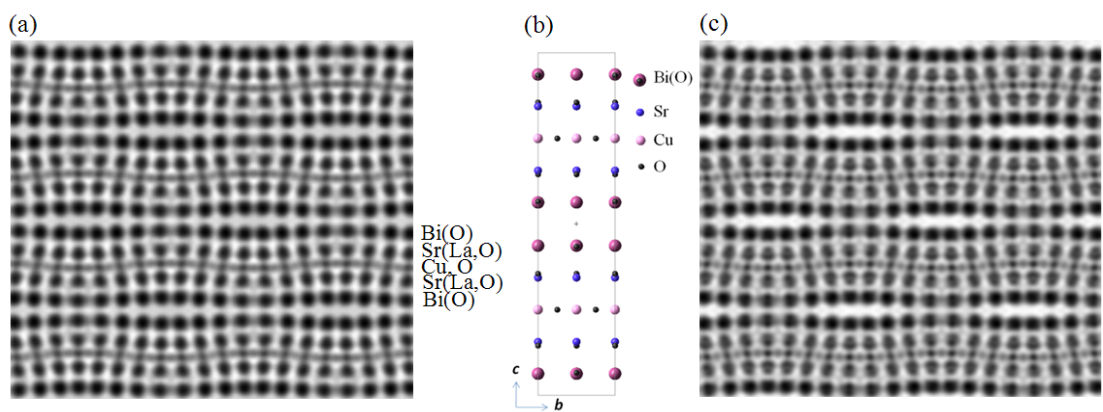

Fig. 2 (a) Symmetry-averaged image of IMS corresponding to the circular area in Fig. 1(b), (b) basic structure model of BSCO, and (c) deconvoluted image of (a) with $C_s$ = 0.05 mm and defocus value $C1$ -13 nm.

To correct the image distortion caused by the contrast transfer function trial deconvolution was carried out with the defocus values in the range from -5 nm to -20 nm with interval 1 nm, and the most possible one is determined to be -13 nm according to the basic structure model of BSCO shown in Fig. 2(b). In the deconvoluted image (see Fig. 2(c)), which represents the projected potential of IMS and hence also is named projected potential map (PPM), Cu and O atoms in $CuO_2$ plane have been resolved, and displacive modulations wave in Cu and O positions can be observed.

Besides PCSI mode with negative focus value, NCSI [19, 23-26] mode with the positive focus value has been widely applied especially to investigate light atoms, such as oxygen in oxides, due to its enhanced image contrast, which is thought to originate from the constructive superposition of linear and non-linear contribution to the total image contrast [24, 32]. In this work, NCSI was also utilized to study the IMS in BSLCO and these two imaging modes were compared through experimental and simulated images. Fig. 3 is the experimental HR image of BSLCO with $C_s$ about -0.013 mm and $C1$ 4 nm (also near Scherzer focus condition). Fourier filtering, HD symmetry averaging and image deconvolution were also carried out for Fig. 3 as mentioned above, and the symmetry-averaged image and deconvoluted one with the defocus value 4 nm are shown in Figs. 4(a) and (b), respectively. Although the atomic number of Bi is higher than that of Sr, image contrast at Bi position is found to be abnormally weaker than that at Sr position in Figs. 4 (a) and (b), which can be explained based on the rule of image contrast change with crystal thickness predicted by the theory of the pseudo-weak-phase object approximation [33]. Same results as to IMS can be obtained from Fig. 4(b) with that from Fig. 2(c) that there are strongly displacive modulations in Bi(O), Sr(La, O) and CuO layers as well as obvious compositional modulations in Bi(O), and Sr(La, O) layers in the examined BSLCO.

Comparing the two PPMs in different imaging mode, obvious difference was noticed that size of dots for heavy atoms in Fig. 2(c) (PCSI) is much larger than that in Fig. 4(b) (NCSI) such as Bi(O) and Sr(La, O) columns. In order to interpret this phenomena, multislice simulation was then carried out based on the basic structure model of BSCO [31].

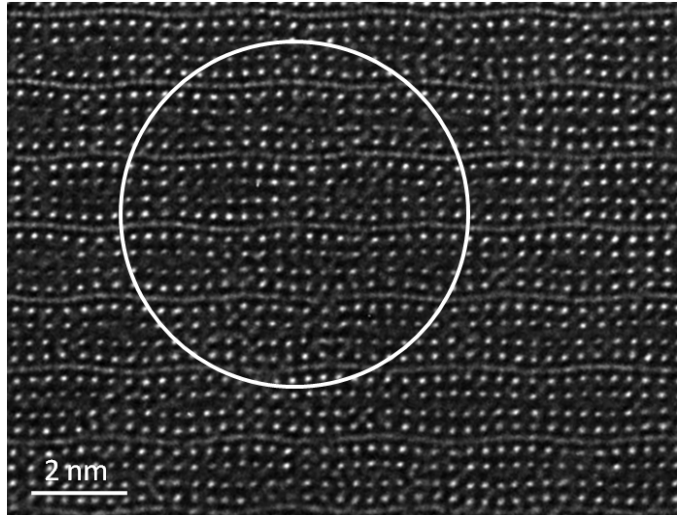

Fig. 3 [100] HR image of BSLCO with $C_s$ = -0.013 mm.

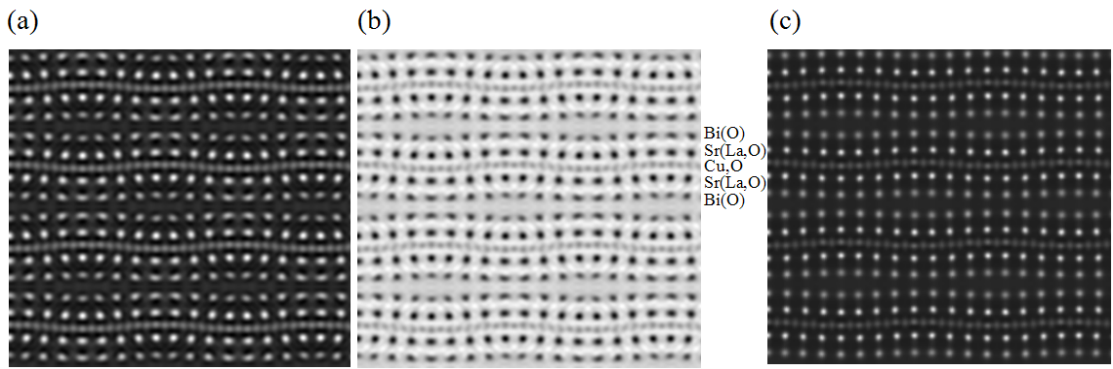

Figs. 4 (a) and (b) Symmetry-averaged and deconvoluted images of BSLCO corresponding to the circular area in Fig. 3, respectively. (c) Simulated image based on the modulation functions with parameters shown in Table 1.

Figs. 5(a) and (b) are the thickness-series simulated image in NCSI mode ($C_s$ = -0.013 mm) and PCSI mode ($C_s$ = 0.013 mm), respectively, with corresponding Scherzer focus conditions. Both of them can be regarded as structural images, but what is different is that in Fig. 5(a) (NCSI mode) white dots represent atoms while the in Fig. 5(b) (PCSI mode) black dots represent atoms. Same as the results obtained from the experimental images, the size of dots in Fig. 5(a), NCSI, is smaller than that in Fig. 5(b), PCSI, especially for heavy atoms Bi(O) and Sr(O) in the examined BSCO. Moreover, with increase of sample thickness the size of dots decreases for NCSI mode, while it increases for PCSI mode. As reported in Ref. [34] object size is one of important factors influencing the precision of the atomic position besides the atomic distance, resolution of the instruments and the number of electron counts, so NCSI mode is more suitable for quantitive structural determination. In the following, the PPM obtained from NCSI image (see Fig. 4(b)) will be used to study IMS in BSLCO with the help of HD space description. Further study about the size of dots in structural images will be given in the other article.

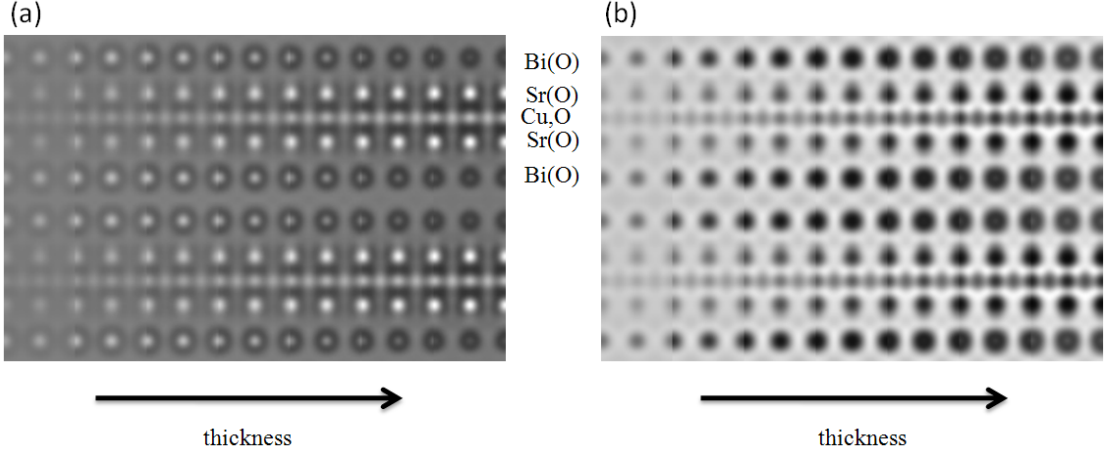

Figs. 5(a) and (b) Simulated image based on the basic structure model of BSCO in NCSI mode ($C$s -0.013 mm, defocus 6 nm) and PCSI mode ($C$s 0.013 mm, defocus -6 nm), respectively. In (a) bright dots represent atoms while in (b) black dots. Thickness for simulation increases from left (0.5 nm) to right (3.8 nm) with an interval 0.55 nm.

## 4.2 Determination of modulated functions

As mentioned in Section 2, atoms for 1D IMS can be treated as one-dimensional and periodic string along the fourth axis in 4D space. So to determinate IMS quantitively, modulation functions should be determined to describe the fluctuation of the string, which in fact is deviation of atomic position and occupation from averaged values, and the method to determine modulated functions with only the limited-size PPM of IMS needed, just like Fig. 4(b), will be introduced briefly [35].

In the case of BSLCO, the modulation vector lies in the ($b$, $c$) plane, so there is discrepancy between the IMS and the average structure (with only main reflection included) in atomic position along the $y$ and $z$ directions and atomic occupancy. A series of deviations as to atomic position denoted by $\Delta y$ and $\Delta z$ and atomic occupancy denoted by $\Delta p$ can be obtained for a certain symmetry-independent atom in different unit cells by comparing the image of average structure and the IMS. According to the HD space description [27], the deviation can be parameterized with the averaged 4D coordinate $t$, which can be determined from the equation

$$t = \mathbf{q} \cdot \bar{\mathbf{r}},$$

where $q$ is the modulation vector, $\bar{r}$ are the coordinates of an atom in the different unit cell in the image of average structure. After series of mathematical processing on $\Delta y(t)$, $\Delta z(t)$ and $\Delta p(t)$, the coefficients of modulation functions $A_j$ and $B_j$ can be derived from the formula

$$\Delta f(t) = \sum_{j=1}^{n}[A_j \cos(2\pi t * j) + B_j \sin(2\pi t * j)].$$

Because the coefficients $A_j$ and $B_j$ steeply decrease with increase of the order $j$, only the first two terms are meaningful.

Based on this method, the modulation functions for all unoverlapped symmetry-independent atoms in BSLCO has been determined with the parameters list in Table 1. For the image contrast is affected by the background subtraction and intensity scaling such that it is hard to precisely measure the occupancy of the atoms, the relative deviation of occupancy is more concerned here. Hence, the average occupancy in Bi position by Bi atoms is set to be 1 although there should be mixture between Bi and Sr(La) atoms considering there exists compositional modulations in Bi layer, and it is same with that in Sr(La) positions, Moreover, occupational modulations in Cu and O sites are neglected for their small atomic numbers and consequently possible much error. HD multislice simulation [36] was carried out with the parameters of the modulated functions shown in Table 1, and the simulated image shown in Fig. 4(c) match the Fig. 4(a) very well with $Cs$ = -0.013 mm, defocus value 4 nm and thickness 6 nm, verifying the validity of the modulated functions.

Table1 Parameters of modulation functions of IMS in BSLCO obtained from the [100] PPM shown in Fig. 4(b) and used to obtain the simulated image shown in Fig. 4(c)

| Symmetry independent atoms | Average position and average occupancy | | Deviations of atomic position and occupancy | Parameters of modulated function | | | |
|---|---|---|---|---|---|---|---|
| | | | | Coefficient ($j$ = 1,2) | | | |
| | | | | $A_1$ | $B_1$ | $A_2$ | $B_2$ |
| Bi | $\bar{y}$ | -0.0015 | $\Delta y$ | 0.0235 | -0.0512 | -0.0035 | 0.0015 |
| | $\bar{z}$ | 0.0642 | $\Delta z$ | 0.0047 | 0.0074 | -0.0036 | -0.0004 |
| | $\bar{p}$ | 1.0000 | $\Delta p$ | -0.0994 | -0.0517 | -0.0398 | -0.0153 |
| Sr | $\bar{y}$ | 0.0008 | $\Delta y$ | 0.0314 | -0.0163 | -0.0044 | -0.0017 |
| | $\bar{z}$ | 0.1714 | $\Delta z$ | 0.0047 | 0.0083 | 0.0000 | -0.0001 |
| | $\bar{p}$ | 0.8000 | $\Delta p$ | 0.0604 | 0.0414 | -0.0029 | 0.0161 |
| La | $\bar{y}$ | 0.0023 | $\Delta y$ | 0.0314 | -0.0163 | -0.0044 | -0.0017 |

|     | $\bar{z}$ | 0.1711 | Δz | 0.0047  | 0.0083 | 0.0000  | -0.0001 |
|-----|-----------|--------|----|---------|--------|---------|---------|
|     | $\bar{p}$ | 0.2000 | Δp | 0.0604  | 0.0414 | -0.0029 | 0.0161  |
| Cu  | $\bar{y}$ | 0.0000 | Δy | -0.0052 | 0.0025 | -0.0024 | 0.0038  |
|     | $\bar{z}$ | 0.2500 | Δz | -0.0025 | 0.0144 | -0.0011 | 0.0003  |
|     | $\bar{p}$ | 1.0000 | Δp | 0       | 0      | 0       | 0       |
| O   | $\bar{y}$ | 0.2531 | Δy | -0.0028 | 0.0050 | 0.0019  | 0.0083  |
|     | $\bar{z}$ | 0.2499 | Δz | 0.0006  | 0.0100 | -0.0001 | 0.0004  |
|     | $\bar{p}$ | 1.0000 | Δp | 0       | 0      | 0       | 0       |

$\bar{y}$ and $\bar{z}$ are average coordinates on the *b* and *c* axis, respectively, $\bar{p}$ is average occupancy. Δz and Δy are deviations of atomic position from average position along *b* and *c* axis, respectively, and Δp is deviations of occupancy from average occupancy. Deviations are expressed in form of Fourier series $\Delta f(t) = \sum_{j=1}^{2}[A_j \cos(2\pi t * j) + B_j \sin(2\pi t * j)]$

## 5 Conclusions

IMS in BSLCO has been studied by means of aberration-corrected electron microscopy in combination with HD space description. PPM of IMS in BSLCO can be obtained with all unoverlapped atoms resolved after image deconvolution, and the displacive modulation in Bi(O), Sr(La, O) and Cu-O layers were found as well as the occupational modulation in Bi(O) and Sr(La, O) layers. Through comparing the PPMs obtained from the images in two different imaging modes, size of dots in NCSI PPM is found to be smaller than that in PCSI one especially for heavy atoms, which is evidenced by image simulation. For size of dots influences the precision of atomic position, the PPM obtained from the NCSI image is used to determine the modulation functions of all unoverlapped atoms with the help of HD space description, and the validation of the functions is verified by HD multislice simulation.

## Acknowledgements

This work was supported by the National Basic Research Program of China (grant number: 2011CBA001001) and the National Natural Science Foundation of China (grant number: 50672124, 10874207 and 11104327). This work made use of the resources of the Beijing National Center for Electron Microscopy at Tsinghua University